\documentclass[amsmath, floatfix, twocolumn, reprint, superscriptaddress, prb, aps, nofootinbib, nobibnotes, longbibliography, final]{revtex4-2} 
\renewcommand{\thispagestyle}[1]{}

\usepackage[T1]{fontenc}
\usepackage[utf8]{inputenc}
\usepackage{graphicx}
\usepackage{latexsym}
\usepackage{amsthm}
\usepackage[low-sup]{subdepth}

\usepackage{color}
\usepackage{mathtools}

\newcommand{\matrixelinl}[3]{\langle #1 \rvert #2 \lvert #3 \rangle}

\usepackage[dvipsnames]{xcolor}
\definecolor{cbred}{HTML}{e31a1c}
\definecolor{cbgreen}{HTML}{33a02c}
\definecolor{cbblue}{HTML}{176aa7}
\definecolor{cborange}{HTML}{ff7f00}
\definecolor{cbviolet}{HTML}{6a3d9a}

\usepackage[]{newtxtext} 
\usepackage[subscriptcorrection,nosymbolsc,smallerops,bigdelims]{newtxmath} 
\DeclareMathAlphabet{\mathcal}{OMS}{cmsy}{m}{n} 
\DeclareMathAlphabet{\mathbcal}{OMS}{cmsy}{b}{n} 
\usepackage{bm}

\usepackage{floatrow}
\usepackage{siunitx}
\mathchardef\mhyphen="2D
\usepackage{hyperref}
\hypersetup{
	colorlinks,
	linkcolor={blue!90!black},
	citecolor={blue!90!black},
	urlcolor=	{blue!90!black}
}

\let\oldeqref\eqref
\renewcommand*{\eqref}[1]{%
	\hyperref[#1]{\oldeqref{#1}}%
}

\usepackage{dcolumn}
\newcolumntype{d}[1]{D{.}{.}{#1}}

\newcommand{\figref}[1]{Fig.~\ref{#1}}
\newcommand{\subfigref}[2]{Fig.~\hyperref[#1]{\ref*{#1}(#2)}}
\newcommand{\subfigsref}[3]{Figs.~\hyperref[fig:#1]{\ref*{fig:#1}(#2)}-\hyperref[fig:#1]{\ref*{fig:#1}(#3)}}

\newcommand*{\newmg}[1]{#1}
\newcommand*{\newms}[1]{#1}
\newcommand*{\newph}[1]{#1}
\newcommand*{\newgs}[1]{#1}

\newcommand*{\kp}{\bm{k}{\cdot}\bm{p}}
\newcommand*{\mr}[1]{\mathrm{#1}}

\begin{document}

\title{Optical and electronic properties of symmetric InAs/InGaAlAs/InP quantum dots formed by a ripening process in molecular beam epitaxy: a promising system for broad-range single-photon telecom emitters}

\author{P. Holewa}
\email{pawel.holewa@pwr.edu.pl}
\affiliation{%
Laboratory for Optical Spectroscopy of Nanostructures, %
Department of Experimental Physics, %
Faculty of Fundamental Problems of Technology, %
Wroc\l{}aw University of Science and Technology, %
Wybrze\.ze Wyspia\'nskiego 27, %
50-370 Wroc\l{}aw, %
Poland%
}

\author{M. Gawe{\l}czyk}
\altaffiliation[Currently at ]{Institute of Physics, Faculty of Physics, Astronomy and Informatics, Nicolaus Copernicus University, 87-100 Toru{\'n}, Poland}
\affiliation{Department of Theoretical Physics, Faculty of Fundamental Problems of Technology, Wroc\l{}aw University of Science and Technology, 50-370 Wroc\l{}aw, Poland}
\affiliation{%
Laboratory for Optical Spectroscopy of Nanostructures, %
Department of Experimental Physics, %
Faculty of Fundamental Problems of Technology, %
Wroc\l{}aw University of Science and Technology, %
Wybrze\.ze Wyspia\'nskiego 27, %
50-370 Wroc\l{}aw, %
Poland%
}

\author{A. Mary\'nski}
\affiliation{%
Laboratory for Optical Spectroscopy of Nanostructures, %
Department of Experimental Physics, %
Faculty of Fundamental Problems of Technology, %
Wroc\l{}aw University of Science and Technology, %
Wybrze\.ze Wyspia\'nskiego 27, %
50-370 Wroc\l{}aw, %
Poland%
}

\author{P. Wyborski}
\affiliation{%
Laboratory for Optical Spectroscopy of Nanostructures, %
Department of Experimental Physics, %
Faculty of Fundamental Problems of Technology, %
Wroc\l{}aw University of Science and Technology, %
Wybrze\.ze Wyspia\'nskiego 27, %
50-370 Wroc\l{}aw, %
Poland%
}

\author{J.~P. Reithmaier}
\affiliation{
 Institute of Nanostructure Technologies and Analytics (INA), Center for Interdisciplinary Nanostructure Science and Technology (CINSaT), University of Kassel, Heinrich-Plett-Str.~40, 34132 Kassel, Germany
}

\author{G. S\k{e}k}
\affiliation{%
Laboratory for Optical Spectroscopy of Nanostructures, %
Department of Experimental Physics, %
Faculty of Fundamental Problems of Technology, %
Wroc\l{}aw University of Science and Technology, %
Wybrze\.ze Wyspia\'nskiego 27, %
50-370 Wroc\l{}aw, %
Poland%
}

\author{M. Benyoucef} \email{m.benyoucef@physik.uni-kassel.de}
\affiliation{
 Institute of Nanostructure Technologies and Analytics (INA), Center for Interdisciplinary Nanostructure Science and Technology (CINSaT), University of Kassel, Heinrich-Plett-Str.~40, 34132 Kassel, Germany
}%

\author{M. Syperek}
\affiliation{%
Laboratory for Optical Spectroscopy of Nanostructures, %
Department of Experimental Physics, %
Faculty of Fundamental Problems of Technology, %
Wroc\l{}aw University of Science and Technology, %
Wybrze\.ze Wyspia\'nskiego 27, %
50-370 Wroc\l{}aw, %
Poland%
}

\begin{abstract}
We present a detailed experimental optical study supported by theoretical modeling of InAs quantum dots (QDs) embedded in an InAlGaAs barrier lattice-matched to InP(001) grown with the use of a ripening step in molecular beam epitaxy. The method leads to the growth of in-plane symmetric QDs of low surface density, characterized by a multimodal size distribution resulting in a spectrally broad emission in the range of 1.4--2.0~\si{\micro}m, essential for many near-infrared photonic applications. We find that, in contrast to the InAs/InP system, the multimodal distribution results here from a two-monolayer difference in QD height between consecutive families of dots. This may stem from the long-range ordering in the quaternary barrier alloy that stabilizes QD nucleation. \newmg{Measuring the photoluminescence (PL) lifetime of the spectrally broad emission, we find a nearly dispersionless values of $1.3\pm0.3$~ns}. Finally, we examine the \newmg{temperature dependence of emission characteristics}. We underline the impact of localized states in the wetting layer playing the role of carrier reservoir during thermal carrier redistribution. \newms{We determine the hole escape to the InAlGaAs barrier to be a primary PL quenching mechanism in these QDs}.
\end{abstract}

\maketitle

\section{Introduction}

Self-assembled InAs quantum dots (QDs) embedded between barriers lattice-matched to InP remain attractive candidates for photon emitters in the near-infrared spectral range \cite{Senellart2017,Buckley2012}. 
It is mostly due to vast possibilities to shape their properties and photonic environment to target specific applications ranging from low-threshold lasers to non-classical photon sources for quantum communication protocols. \newms{The latter is especially difficult to achieve, since the typical growth of InAs QDs on the InP substrate by moleular beam epitaxy (MBE) leads to formation of strongly in-plane asymmetric objects of a significant areal density reaching $10^{10}$--$10^{11}$~cm$^{-2}$. Such high surface coverage prevents sufficient spatial and spectral isolation of individual QDs, while large QD sizes result in an atypically rich and dense optical spectrum unfavorable for single-dot applications \cite{Gawelczyk2018,Gawelczyk2019}}. The growth process of InAs QDs on an InP(001) substrate and control of their parameters are still challenging and comprise many technological steps and details \cite{Khan2014}.

Recently, it has been proposed \cite{Benyoucef2013, Yacob2014} to utilize an additional growth step in MBE of InAs QDs on InP(001) that mimics the Ostwald ripening known for the formation of colloidal micro-crystals. 
During the ripening process initially formed QDs decompose, and the material is redistributed between other dots, which typically leads to splitting of the initial QDs size distribution into distinct families. The ripening technique has been already applied to other semiconductor QD systems like Ge/Si \cite{Ross1998}, PbSe/PbTe \cite{Raab2000} or InAs/GaAs \cite{Poetschke2004}. In the case of InAs/InAlGaAs/InP(001) QDs, it has been shown \cite{Yacob2014} that the mean size of the dots depends on the ripening temperature. These ripening-assisted grown InAs QDs on InP substrate are particularly promising for quantum information processing as their recent development led to demonstrations of single-photon emission at $\sim1.45$~\si{\micro}m \cite{Benyoucef2013} \newms{(the S band)}, coupling to optical modes of photonic crystal microcavities \cite{Kors2017}, and triggered single-photon emission with high purity \cite{Musial2019}.

\newms{Although some optical experiments on these QDs have already been presented \cite{Benyoucef2013,Yacob2014,Kors2017,Kors2018,Musial2019}, we would like to elaborate here on possible further benefits that come from a single-stage ripening process by pointing at their so-far unexplored and not described application-relevant properties.
We focus on: (a)~broad spectral coverage of QD emission superimposed with the S ($\SIrange{1460}{1530}{\nano\meter}$), C ($\SIrange{1530}{1565}{\nano\meter}$), L ($\SIrange{1565}{1625}{\nano\meter}$), and U ($\SIrange{1625}{1675}{\nano\meter}$) transmission bands of silica fibers, what can be useful for the multiband quantum-secured transmission protocols highly sought for overcrowded long-haul optical networks; (b)~the multimodal QD size distribution allowing simultaneously for better spectral isolation and filtering of single-dot emission, as well as for assuring spatial isolation of a spectrally chosen QD for site-selective engineering of the photonic environment \cite{Sapienza2015} to control the QD emission properties in a future device; (c)~thermal stability of emission from QDs, important for applications in devices operating at elevated temperatures.}

\newgs{The flexibility in the choice of a quantum emitter from a single wafer in a very broad spectral range covering all the high transmission telecommunication bands can be of practical relevance for certain future applications. It especially concerns all the concepts based on wavelength division multiplexing (WDM) allowing to increase the total transmission rate through the fibers or the optical system efficiency and functionality in analogy to such schemes known from classical communication. For instance, there exist both predictions and experimental demonstrations of the so called multi-user quantum key distribution employing WDM for quantum networks \cite{Brassard2003} and showing recently a really impressive increase in the achievable secure key rates \cite{Eriksson2019} and even a practical implementation in
a node–free eight-user metropolitan quantum communication network \cite{Joshi2020}. WDM is also utilized in the on-chip operation using quantum light in photonic integrated circuits \cite{Elshaari2017} for linear optics quantum computation, where the telecom wavelength range is demanded to combine the III-V single photon emitters with silicon technological platform. Another branch is gas spectroscopy or LIDAR-like sensing systems single photon emitters in the near infrared  considered as prospective and competitive in imaging of gas leaks and dangerous vapors or to perform quantified remote mapping of gases to monitor air quality in agricultural and metropolitan areas \cite{idquantique}. Here, the broad range selection of the quantum emitters is also profitable to allow detection of many environmentally relevant gasses, like e.g. CO$_2$, N$_2$O, H$_2$O, CH$_4$, having their strong absorption lines in the 1400--1650~nm range. In that context also more sophisticated single-photon-based gas detection solutions employing WDM or on-chip spectrometers for, e.g., astronomical observation, spectroscopic imaging and quantum communications, are also reported \cite{Cheng2019}.}

\newph{Here, we report on the optical investigations of InAs/InAlGaAs/InP QDs, which we support by theoretical modeling.
We observe the multimodal distribution of QD sizes and explain the emission-energy separation of consecutive QD families by their modeling. With this, we obtain a strong indication that their heights differ by 2~monolayers (MLs) of InAs.
This in turn suggests the presence of long-range ordering of atoms in the barrier alloy, what may enhance the reliability of future photonic devices, as it reduces the crystal-lattice free energy \cite{Stringfellow1991,Stringfellow1989,Moeck2005}, stabilizing the entire strained QD system.}

The detailed goals of this work are to: (i) find the links between QD parameters (height, in-plane size, and chemical composition) and the resulting broad spectral range and other characteristics of emission, (ii) study the character and importance of WL states, (iii) assess the impact of exciton confinement in a QD on recombination dynamics, (iv) investigate the thermal carrier redistribution processes and optical response of the dots at elevated temperatures. These objectives are achieved by using several complementary spectroscopic tools: modulated reflectivity ($\Delta R/R$), excitation-power-, polarization-, and temperature-dependent photoluminescence (PL), and time-resolved photoluminescence (TRPL), all supported by numerical modeling of exciton states in QDs and WL within the multi-band envelope-function $\kp$ theory. The collected data and conclusions drawn can trigger the work on such ripening-assisted-grown QDs towards unveiling their physical and chemical properties and their use in photonic applications in the near infrared.

\section{Experimental and theoretical methodology}
In this section we provide details about sample growth, experimental setups, and theoretical framework used to study optical and electronic properties of the structure.

\subsection{Sample growth}
The investigated sample was grown on (001)-oriented InP substrate using an MBE reactor equipped with two valved solid-source arsenic and phosphorus cracker cells. The growth sequence starts from a 100-nm-thick InP buffer layer directly deposited on the substrate at 465$^\circ$C and followed by a nominally 228-nm-thick In$_{0.53}$Al$_{0.24}$Ga$_{0.23}$As layer lattice matched to InP. A growth interruption lasting for 35~s was introduced to allow for the group-V atoms exchange on the InP/InAlGaAs interface. Subsequently, the Stranski-Krastanov (S-K) growth mode was used to form an initial layer of QDs by deposition of nominally 2~ML of InAs at 514$^\circ$C with a growth rate of 0.4~ML/s. Following the S-K QD formation, the ripening process starts during the reduction of substrate temperature down to 413$^\circ$C with a cooling rate of 30~K/min. under As$_{2}$ pressure of $6\times 10^{-6}$ Torr \cite{Yacob2014}. After the cooling process, the dots were capped by a nominally 228-nm-thick In$_{0.53}$Al$_{0.24}$Ga$_{0.23}$As layer. During the growth of the cap layer, the substrate temperature was kept at 413$^\circ$C for the first 20~nm. Afterwards, it was increased with a rate of 30~K/min up to 514$^\circ$C and then kept constant for the remaining layer thickness. More details about the growth procedure can be found elsewhere \cite{Benyoucef2013, Yacob2014}.

\begin{figure*}[t!]
\includegraphics[width=\linewidth]{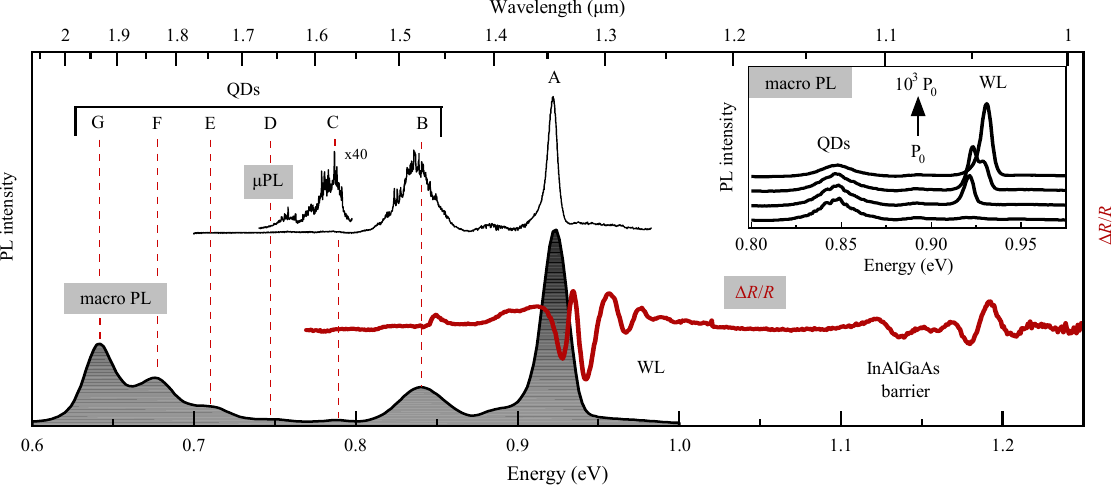}
\caption{\label{fig:PL_MR}(Color online)
Photoluminescence (macro PL, grey shaded area), high-resolution PL (\si{\micro}PL, solid black line), and modulated reflectivity ($\Delta R/R$, solid red line) from the structure with QDs at $T=10$~K. WL is the wetting layer. QDs-related PL emission bands are labelled as B--G, where B denotes remnants of initially grown S-K QDs and maxima C--G represent ripening-assisted-grown QDs. Inset: Power series of macro PL from the B family of QDs and WL.}
\end{figure*}

\subsection{\newmg{Experimental details}}
For spectroscopic experiments, the structure was held in a helium closed-cycle refrigerator allowing for control of the sample temperature in the range of $\SIrange{10}{300}{\kelvin}$. In the case of PL and TRPL experiments, the sample was excited by a train of $\sim2$-ps-long pulses with $\sim13.2$~ns pulse-to-pulse interval and $\sim1.48$~eV photon energy. \newms{The laser spot was focused to $\sim\SI{150}{\micro\meter}$ in diameter}. Emission from the structure was collected in a standard far-field optical setup and dispersed by a 0.3~m-focal-length monochromator. Time-integrated PL spectra were measured in a wide spectral range of $\SIrange{1.2}{2.1}{\micro\meter}$ via the lock-in technique at the reference modulation frequency of 2~kHz, using a thermoelectrically cooled InAs-based single-channel detector. The HR-PL was measured in the same setup. However, in this case the excitation and collection of emission were performed through a microscope objective with high numerical aperture $\mathrm{NA}=0.65$. \newms{The laser spot was defined to $\sim\SI{2}{\micro\meter}$ in diameter}. A liquid-nitrogen-cooled InGaAs-based linear camera detector registered the HR-PL spectrum in the limited spectral range up to 1.65~\si{\micro}m. The TRPL was measured by a time-correlated single photon counting method. Photons were spectrally filtered by a monochromator and subsequently collected by the NbN superconducting detector. The multi-channel event timer was synchronized to the pulse train to produce photon event statistics. The overall temporal resolution of the TRPL setup was $\sim80$~ps.

In the $\Delta R/R$ experiment, a halogen lamp was used as a broadband probe beam source. The 630~nm line from a semiconductor diode laser was employed for photomodulation purposes. The 0.3-m-focal-length monochromator dispersed the white light reflected from the sample normalized changes of which were measured via the lock-in technique similarly as the PL. 

\subsection{\newmg{Theoretical framework}}
Initial calculations of single-particle states in the WL and QDs were performed with the commercially available \textit{nextnano} software \cite{nextnano,Birner2007}, which utilizes the continuum elasticity model for the strain distribution, and afterwards calculates the electron eigenstates within the 8-band $\kp$ method including the strain-driven piezoelectric field. Relevant material parameters were taken from Ref.~\cite{Vurgaftman2001}. The calculations allowed for establishing the bounds for QD parameters (height, lateral dimension, and chemical content) leading to the multimodal energy distribution, analysis of the WL states, and separation between relevant energy states in the investigated structure as an input for discussion on the PL quenching mechanisms.

This was followed by extended band-structure calculations within the state-of-the-art implementation of the multiband $\kp$ method \cite{Gawarecki2014,Gawarecki2018}, including spin-orbit interactions, structural strain, and piezoelectric field up to second-order terms in the strain tensor.
Next, exciton states were found within the configuration-interaction approach by diagonalizing the Coulomb and phenomenological electron-hole exchange interactions in the basis of $32\times 32$ electron-hole state configurations, where hole states were obtained by time reversal of valence electron ones. \newmg{The number of single-particle states included in the calculations was determined on the basis of convergence tests.}
Then, oscillator strengths and radiative lifetimes for exciton states were obtained within the dipole approximation \newmg{by computation of matrix elements of the momentum operator $\bm{P}=(m_0/\hbar)(\partial H/\partial \bm{k})$, where $m_0$ is the electron mass.
The oscillator strength for the  $i$th exciton state composed of a number of electron-hole configurations (single-particle states are labeled by $\alpha$ and $\beta$) is then $f_i = 2 / (m_0 E_i) \lvert\sum_{\alpha\beta} c_{\alpha\beta}\matrixelinl{\bm{\psi}_{\mathrm{v}}^{(\alpha)}} {\bm{P}} {\bm{\psi}_{\mathrm{c}}^{(\beta)}} \rvert^2$.
Finally, we calculated radiative lifetimes, connected with oscillator strength $f$ by the relation that may be approximately written as $\tau[\si{\nano\second}]=45 (\lambda [\si{\micro\meter}])^2 /nf$, where $\lambda$ is the emission wavelength, and $n$ is the refractive index.
The derivation may be found in Ref.~\cite{Karrai2003}, while the details of modeling and parameter taken for the InAs/InAlGaAs/InP material system may be found in Ref.~\cite{Gawelczyk2017}}.

\section{Results and discussion}
In this section, we show experimental results supported by theoretical calculations and provide discussion on the properties of the investigated QD structure.

\subsection{Modulated reflectivity and photoluminescence}
The measured modulated reflectivity spectrum is plotted with a solid red line in \figref{fig:PL_MR} and shows two noticeable $\Delta R/R$ features settled in the ranges of 1.13--1.20~eV and 0.92--0.96~eV, respectively. While the former is attributed to optical absorption involving valence and conduction bands of the InAlGaAs barrier \cite{Syperek2018}, the latter is tentatively assigned to absorption in the WL. The WL-related feature appears to be constituted by at least two transitions centered at $\sim$0.93~eV and $\sim$0.94~eV. The PL spectrum presented as a grey-shaded area in \figref{fig:PL_MR} helps to reveal the nature of both these features. It shows an intensive emission band labeled as A and centered at about $\sim$0.93~eV, which overlaps with the low-energy \newmg{part of the} $\Delta R/R$ feature. More interestingly, it changes with increasing optical-pumping power $P$ as shown in the inset to \figref{fig:PL_MR}. For the low pumping power $P=P_{0}$ the band A is centered between 0.92--0.93~eV and shifts towards higher energies with increasing $P$. We tentatively stated that PL band A and low-energy $\Delta R/R$ feature are related to optical transitions among the zero-dimensional (0D) localized states in the WL \cite{Gammon1996, Leosson2000, Robinson2001,Babinski2008,Sek2010a,Ning2011,Syperek2013}, whereas the higher-energy $\Delta R/R$ transition involves its 2D band edges. Consequently, the observed evolution of the PL peak A with $P$ can be related to gradual filling of the WL density of states (DOS) starting from its 0D-like tail (0D DOS) and ending in the 2D states. The 0D DOS may originate from an inhomogeneity of the WL due to fluctuation of its thickness and chemical composition, possibly introduced during the ripening stage of the MBE growth. Alternatively, the WL inhomogeneity could be triggered by chemical fluctuations in the InAlGaAs barrier as suggested by the spectral smearing of the barrier-related $\Delta R/R$ feature.

\begin{figure}[tb]
	\includegraphics[width=\linewidth]{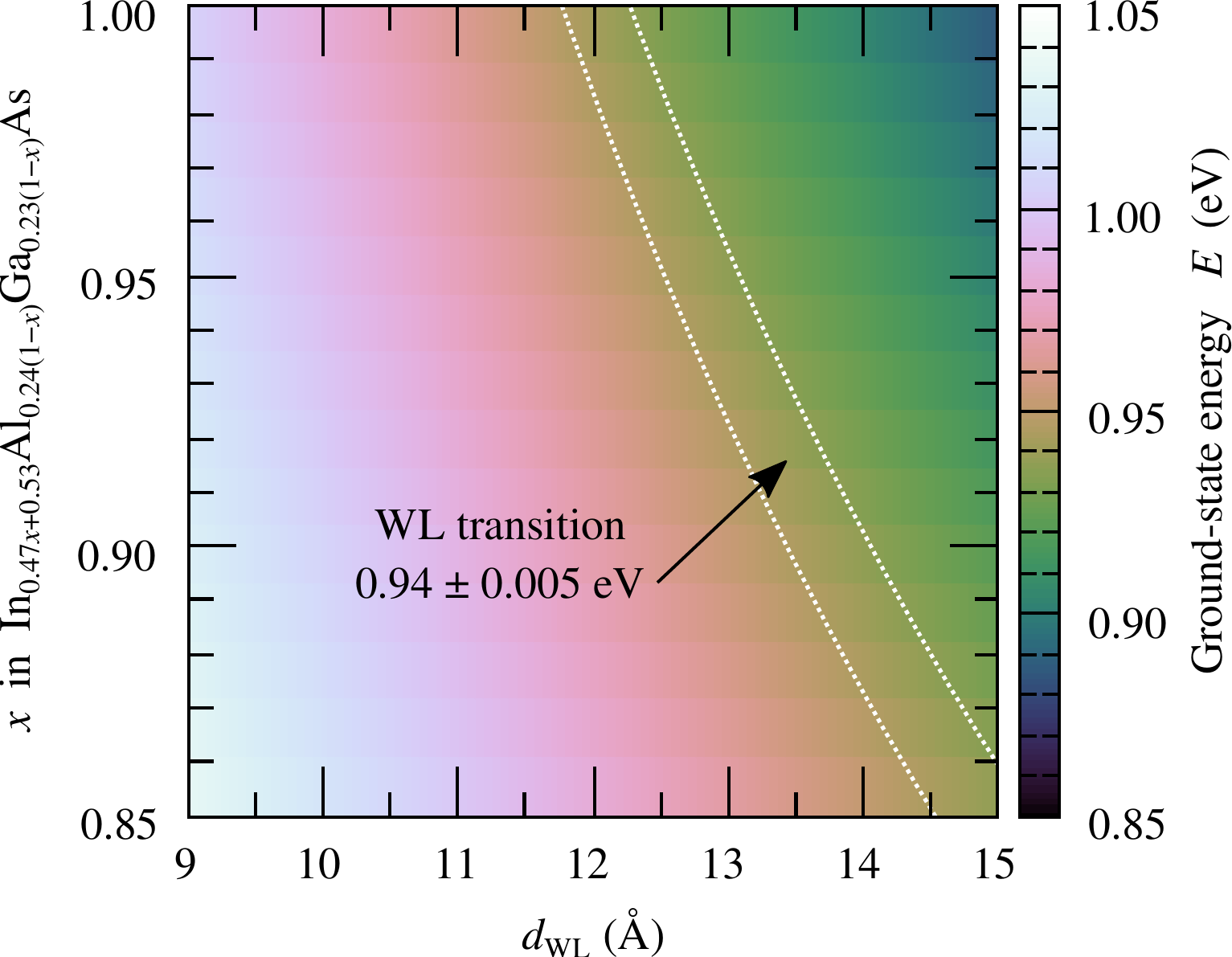}
	\caption{\label{fig:WL}(Color online) \newmg{Color map of calculated wetting layer ground-state energy as a function of WL width $H$ ($x$-axis) and material composition ($y$-axis). Dotted white lines mark the energy of WL transition observed in the $\Delta R/R$ signal.}
	}
\end{figure}
The 2D band edge-related transition in the WL at $\sim$0.94~eV allows \newmg{us to determine the range of possible WL characteristics (composition $x$ of the In$_{0.47x+0.53}$Al$_{0.24(1-x)}$Ga$_{0.23(1-x)}$As alloy and WL thickness $d_{\mr{WL}}$). For this purpose, the electronic structure of the WL was calculated, using the \textit{nextnano} software \cite{nextnano}, for $d_{\mr{WL}}$ varied from 0.9~nm to 1.5~nm and $x$ from 0.85 to 1. The calculated transition energies are presented in \figref{fig:WL} as a color map, where the two dotted lines distinguish the area of energies corresponding to the WL transition. As we do not expect very strong intermixing, this calculation allows us to determine that the average WL characteristics have to be between 4~ML ($\sim1.2$~nm) of pure InAs and 5~ML ($\sim 1.5$~nm) of alloy with $x=0.85$, which resembles the recent transmission electron microscope images of similar QDs \cite{Carmesin2017}. In the latter a rather soft interface with a gradient of composition was found. Based on this, in further calculations done for QDs, we take a WL consisting of 4~MLs of pure InAs and perform Gaussian averaging to simulate the soft interface. Nonetheless, these subtle WL details have in fact only a very minor impact on results of QD calculation presented further.} 

\figref{fig:PL_MR} shows also a series of PL peaks below the WL-assigned optical transitions with their positions extending down to $\sim$0.6~eV. At least six such bands are resolved and labeled from B to G. These are attributed to emission from QDs with multimodal size distribution formed during the ripening step in the MBE growth. The band B has a relatively high intensity as compared to bands from C to G, for which the intensity gradually increases. It is reasonable to assume that QDs from family B are the remnants of initially grown S-K QDs. During the ripening stage, material from partially decomposed S-K dots is transferred to other dots forming the C--G families. This scenario is supported by previously published structural data \cite{Yacob2014}, where the surface density of the remnants of the initially grown QDs (10$^{9}$--10$^{10}$~cm$^{-2}$) can be comparable to the one of newly formed QDs after the ripening stage. Interestingly, for QDs belonging to the C family emitting in the telecom spectral range of 1.53--1.63~$\si{\micro}$m the surface density seems to be the lowest among all families. Indeed, the $\si{\micro}$PL spectrum presented in \figref{fig:PL_MR} obtained from unprocessed sample (without any special patterning of the structure, e.g., mesa etching) shows a discrete spectrum, i.e., sharp and narrow PL lines originating from single QDs. The same is hardly resolvable for QDs in the band B and disappears completely for the band A, suggesting high areal density of the 0D potential traps in the WL, additionally confirmed by its high PL intensity dominating the emission spectrum.

\begin{figure}[tb]
	\includegraphics[width=\linewidth]{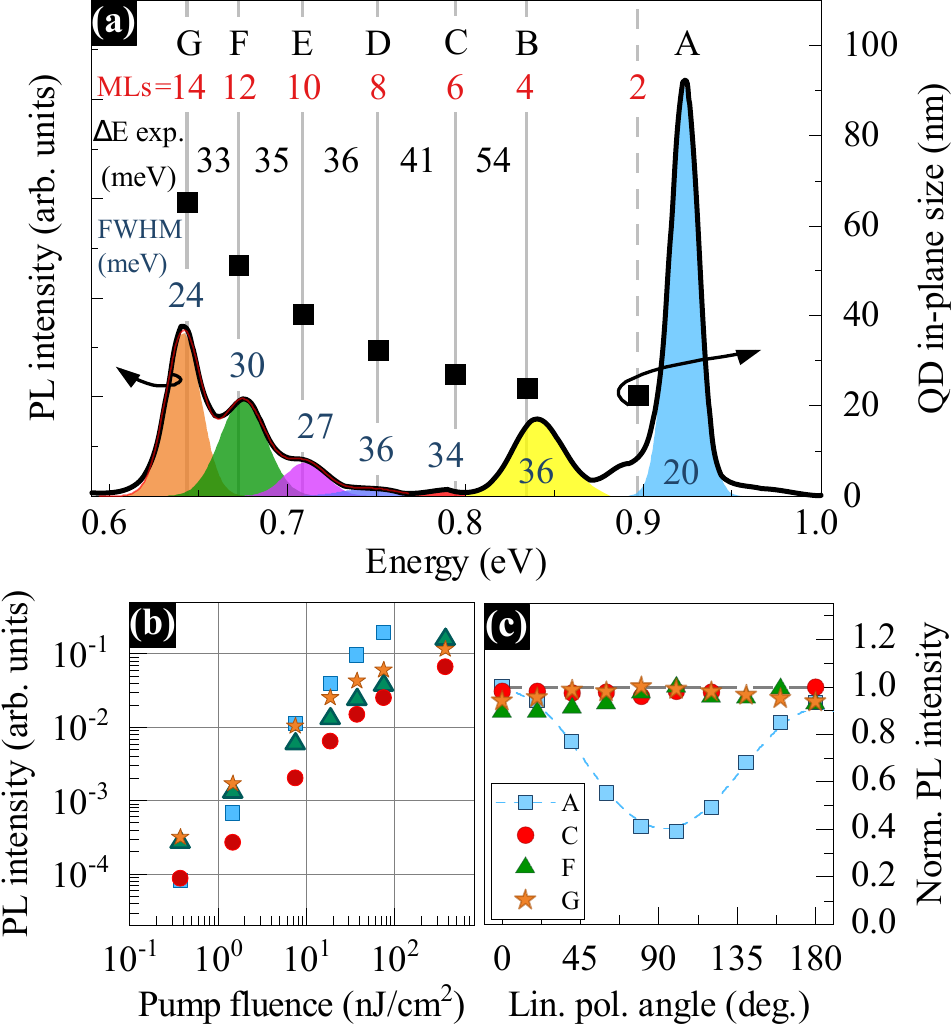}
	\caption{\label{fig:PL_at_12}(Color online)
		(a) The PL spectrum (solid black line) at $T=10$~K with Gaussian fits (color shaded areas for consecutive PL bands). Letters label the peaks, $\Delta$E is the energetic distance between PL bands, the FWHM is full-width-at-half-maximum for the peaks. Vertical grey lines indicate energies obtained from numerical calculations for a given QD in-plane size. Dashed grey vertical line indicates potential remnants of initially grown S-K QDs with 2ML height. Full squares represent the QD in-plane size, determined from the calculations of the ground-state energies.
		(b) Pump-fluence-dependent PL for selected PL bands: A (blue squares), C (red circles), F (green triangles) and G (orange stars). (c) Normalized PL intensity from selected PL bands as in (b), but as a function of the angle of linear polarization in respect to the laboratory frame.}
\end{figure}
The bands B to G originate mostly from ground-state recombination in QDs, rather than from excited states. This conclusion is drawn from the analysis of the pump-fluence-dependent PL presented in \subfigref{fig:PL_at_12}{b}, which shows a nearly linear increase of the peak intensity with pump fluence across almost three decades. In further analysis of PL spectra we extracted the full width at half maximum (FWHM), peak position, and peak-to-peak distance ($\Delta E$) between consecutive bands. These were obtained by fitting each PL peak with a Gaussian function. The relevant parameters are shown at the top of \subfigref{fig:PL_at_12}{a}. The PL bands B to G have the FWHM ranging from 24~meV to 36~meV, which reveals similar inhomogeneity within each of QD families that could be in-plane size fluctuations, strain inhomogeneity, and chemical content variation. We observe an expected super-linear decrease in $\Delta E$ from $\sim54$~meV between peaks B and C down to $\sim33$~meV between G and F resulting from the quantum size effect. This issue will be further addressed in the theoretical considerations below.

In addition, the angle-resolved linear polarization of PL is presented in \subfigref{fig:PL_at_12}{c} for the selected PL bands A, C, F, and G to keep the picture clarity. Surprisingly, an almost 40\% degree of linear polarization (DOLP) is observed for the band A whereas for the QD-related bands C, F, and G the DOLP is negligible within the 5\% of experimental accuracy. The lack of significant DOLP for the QDs can be linked to their high in-plane symmetry \newmg{that for similar QDs was shown to} lead to a negligible fine-structure splitting of the confined exciton states \cite{Kors2017,Musial2019}.
In contrast, high DOLP for the A band may be related to the built-in optical anisotropy in the WL induced by local strain variations and its anisotropy or chemical disorder at InAs/InAlGaAs or InAlGaAs/InP interfaces.

\subsection{\newph{Long-range atom ordering and two-monolayer QD height steps}}
Previous observations of multimodally-distributed InAs QDs embedded directly within the InP barrier were interpreted as resulting from the QD size variation in the form of 1~ML changes in their height between consecutive families \cite{Raymond2003, Tomimoto2007, Gelinas2010}. A similar observation and interpretation were also reported for InAs/GaAs QDs \cite{Poetschke2004,Guffarth2004}. This does not seem to apply to the investigated QDs since the 1~ML ($\sim 0.3$~nm) change in QD height would lead to much smaller energy separation between PL bands than actually observed. Instead, we propose that the height of consecutive QD families changes by 2~ML steps. This hypothesis can find its justification in the expected long-range atom ordering that takes place in the quaternary barrier alloy.
The ordering arises spontaneously for \newph{a multitude of} III-V alloys as the coherent placing of atoms reduces the free energy of the crystal lattice in comparison to a disordered material or to a two-phase alloy \cite{Stringfellow1991}.

The resulting structure has a modulation of composition along particular crystallographic directions where the elemental cell can be expanded as compared to, e.g., a binary alloy. Therefore, if the barrier above the investigated QD layer is organized in the direction [001] into In-rich and In-poor monolayers alternately then the effective height of a QD may be changed only by 2~ML.

Previous observations of long-range ordering include AlGaAs/GaAs \cite{Kuan1985}, InAlAs/InP \cite{Ueda1989,Han1998}, InGaAs/InP(110) \cite{Kuan1987,Shahid1987}, and InAsSb/InSb(001) \cite{Jen1989} hetero-interfaces, \newph{what establishes a background for our conclusion}.
The most commonly arising is CuPt-type superlattice (L1$_1$ structure) resulting in ordering of cation lattices along the [111] direction \cite{Han1998}. Importantly, it was reported for InGaAsP quaternary alloy lattice-matched to InP(001) \cite{Shahid1987,Shahid1988}. 
The CuPt-type structure ordering along the [001] direction relevant for this work was modeled \cite{Ishimaru1995} and observed for, e.g., bulk InAlAs grown on InP(001) substrate \cite{Ueda1989,Kurihara2004}. 
Another, less frequently investigated ordering is the CuAu-I type, which also results in alternating layers along [001] direction reported for MBE-grown strained InGaAs/InAlAs multiple quantum wells on the InP(001) substrate \cite{Lee2005}.
\newph{The atomic ordering was observed also for various III-V and II-VI systems with epitaxial nanostructures \cite{Moeck2001,Moeck2005}.}
\newph{For applications the long-term stability of QDs and their properties is crucial.
The size and composition of a highly strained QD after its growth depend significantly on the conditions in the barrier.
Thermodynamic calculations suggest that QDs in atomically ordered alloys may be more stable than those grown in random lattices \cite{Moeck2005}.
There are also, e.g., \textit{ab initio} total energy \cite{Mbaye1987,Srivastava1985} and strain energy calculations \cite{Moeck2005} showing that ordered structures are more stable than random alloys \cite{Stringfellow1989}.
These results hold for systems with large differences in the lattice constants between the binary constituents, however, the ordering is observed also when this condition is not true. 
Therefore, other models (kinetic \cite{Ishimaru1995} and surface thermodynamic considerations) are proposed to understand this phenomenon for low-mismatch alloys, as in the case of alternating rows of large and small atoms building-in along the growth steps (this is the proposed mechanism for explanation of, e. g., the origin of the ordered CuPt-type structure \cite{Suzuki1988}).
These models help to explain the very first report of the ordering which was observed for the AlGaAs alloy \cite{Kuan1985}, while AlAs and GaAs have very similar lattice constants.}

\newph{Although the long-range ordering in III-V ternary and quaternary alloys has been confirmed by crystallographic methods, it has never been observed for InAlGaAs quaternary alloy investigated here.
Even though our research focuses only on optical properties and does not include the direct investigation of the structural properties, we deduce a similar ordering for InAlGaAs as it has been observed for its constituents \cite{Ueda1989,Han1998,Kuan1987,Shahid1987}, InGaAs and InAlAs, and there is no evident reason that their mixture should organize in a different way.
Additionally, there are some premises, e.g. for GaInAs(P) \cite{Shahid1987}, that ordering in quaternaries may be even more pronounced than in ternaries.}

\subsection{Modeling of QDs}
The observed multimodal QD emission can be explained based on calculations of the QD ground state as a function of QD parameters (height, lateral dimensions, and chemical composition). The modeled QD geometry is a truncated pyramid with a square in the base and the angle between the side facets and the base of 25 degrees, in accordance with the structural studies of QDs grown under similar conditions \cite{Carmesin2017}. \newmg{The dot is settled on a 4-ML-thick WL, and the barrier material is In$_{0.53}$Ga$_{0.23}$Al$_{0.24}$.}

\begin{figure}[tb]
	\includegraphics[width=\linewidth]{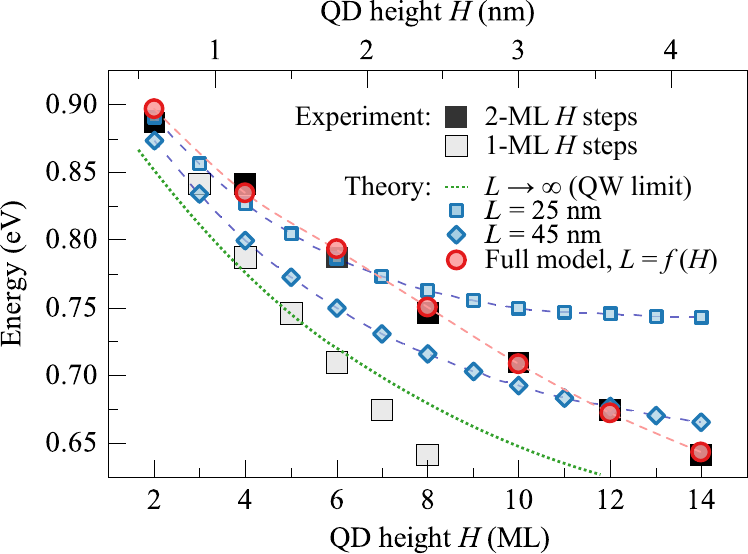}
	\caption{\label{fig:Nextnano_calc}(Color online) \newmg{Comparison of observed QD emission energies, plotted assuming 2-ML (black squares) and 1-ML (gray squares) steps in QD heights, with trends calculated in a simple model for QDs with $L=25$~nm (blue squares) and $L=45$~nm (blue diamonds), as well as in full model including the exciton binding energy and varying dependence $L=f(H)$. The dotted green line shows energy calculated for a quantum well. Height $H$ is measured from the top of WL.}}
\end{figure}
\newmg{We begin with two initial series of single-particle calculations, in which we take almost pure InAs (In$_{0.95}$Al$_{0.025}$Ga$_{0.025}$As) in the WL and a purely InAs QD with height varying by 1~ML and two fixed in-plane sizes, $L=25$~nm and $L=45$~nm, plotted in \figref{fig:Nextnano_calc} with blue squares and diamonds, respectively. These results are compared with the experimental data plotted with black squares at 2-ML height steps and with gray squares at 1-ML steps. While the qualitative agreement is very weak, one may notice that the assumption on 2-ML QD height variation, and thus the presence of long-range ordering and the resulting enhanced stability of QDs, occurs correct here. Additionally, with dotted green line we plot the ground-state energy for the limit of $L\to\infty$, i.e., for a quantum well (QW) as high as the studied QDs. The experimental data plotted assuming 1-ML steps (gray squares) is so steep that it even crosses this line, which makes it nonphysical, as it implies QDs having lower energy that for QW. Thus, at this point we establish that} the G family with emission around 0.65~eV is assigned as having $H=14$~ML, while $H=4$~ML characterizes the B family centered near 0.84~eV. Within this assumption, the smallest dots having 2~ML height should emit near 0.89~eV. Indeed, at this energy the PL spectra in \figref{fig:PL_MR} and \figref{fig:PL_at_12} (vertical dashed grey line) reveal a PL band of low intensity overlapping with tails of bands A and B, which thus comes from the 2~ML-height QDs.
	
For two probe cases related to fixed $L$, the calculated ground-state energies as a function of $H$ deviate from the postulated trend especially for larger heights. The obtained energy change saturates for large $H$, as predicted by the $H^{-2}$ dependence. Therefore, to obtain an agreement between the experimental and theoretical points one needs to introduce an additional, non-linear dependence that relates $H$ with $L$ already presented in \figref{fig:PL_at_12}. \newmg{It allowed us to achieve a good qualitative agreement, the respective theoretical values are plotted with red circles. This final series of calculations was performed in an extended model taking into account the electron-hole Coulomb interaction and soft material interfaces. Starting with} nominally clean InAs inside WL and QDs, Gaussian averaging (with $\sigma = 0.3~\mathrm{nm}$) of the material profile has been used to simulate material intermixing at interfaces. QDs of height from 2 to 14~MLs were simulated, with the in-plane size varied in the range of $L = 25$--70~nm. The resultant dependence of the ground-state energy on these dimensions was very smooth, thus it allowed for extraction of the $L=f(H)$ dependence that was then fit to the experimental values. Finally, this nonlinear dependence introduced into QD modeling resulted in a very good agreement of the calculated energies with PL peaks' positions (see red circles in \figref{fig:Nextnano_calc}).

\newmg{The introduction of variable $L$ might suggest that in principle the experimental data could be also reproduced assuming 1-ML height steps, by using a strong enough dependence of $L$ on $H$. This is, however not the case. As pointed above, the 1-ML-stepped experimental data crosses the theoretical result obtained for a quantum well, which means that the hypothetical $L(H)$ dependence is divergent. Thus, such a sequence of ground-state energies could not be obtained for QDs varying in height by single MLs.}

\subsection{Carrier dynamics}
\begin{figure}[tb]
\includegraphics[width=\linewidth]{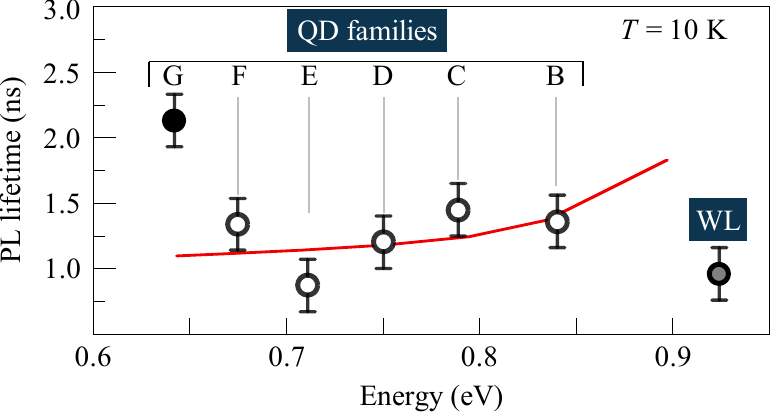}
\caption{\label{fig:PLDecay2}(Color online) PL lifetimes at $T=10$~K for different families of ripening-assisted grown InAs/InAlGaAs/InP(001) QDs (PL bands B--G) and the WL (PL band A). Solid red line in is a calculated trend for the PL lifetime.}
\end{figure}

The theoretical calculation the excitonic states in QDs allowed us also to determine the exciton recombination dynamics presented as a solid red line in \figref{fig:PLDecay2}. The result of the comparison with the experimental values may shed some light onto the correctness of the assumptions made and the legitimacy of using a specific set of QD structural parameters.

The TRPL experiment was performed at $T=10$~K. \newph{The PL decay curves measured at each of the respective PL peaks shown in \subfigref{fig:PL_at_12}{a} were analyzed using the PTI FelixGX software by Photon Technology International. 
Using the maximal entropy method (MEM) \cite{JaynesPR1957, JaynesPR1957a}, the PL lifetimes are extracted without initial knowledge or assumptions on the number of underlying processes. 
Due to this, MEM has proved effective in reconstruction of decays consisting of several exponential components \cite{Livesey1987}.}
\newph{In this approach, one obtains a distribution of times of nonzero width, which results from the incomplete (noisy) information, as long as not really present in the investigated case.
The reconstruction formula is}
\begin{equation}
	I(t) = I(0) \times \sum_{i=1}^N A_i \exp{ \left(-\frac{t}{\tau_i}\right)},
\end{equation}
\newph{where $N$ is the number of components in the quasi-continuous distribution of lifetimes describing the decay, and coefficients $A_i$ are the corresponding amplitudes.
If the distribution is a set of well-resolved peaks, it is valid to treat their means as single representative lifetimes.
Such averages are plotted with circles in \figref{fig:PLDecay2}.}

\newmg{The PL lifetimes obtained for QD families B-F ($1.4\pm0.2$~ns, $1.5\pm0.2$~ns, $1.2\pm0.2$~ns, and $1.1\pm0.2$~ns, respectively) are similar and, as their variation is on the level of the experimental uncertainty, no clear trend can be determined. These overall values that average to $1.3\pm0.3$~ns agree well with those obtained theoretically. In theory, a noticeable increasing trend is present, due to effectively weaker (relative to single-particle level spacing) electron-hole Coulomb interaction in smaller QDs. However, the uncertainty of experimental values does not allow this result to be verified. }
At the emission energy of 0.84~eV, a similar PL lifetime of $\sim 1.65$~ns has been obtained for slightly in-plane asymmetric S-K InAs/InAlGaAs/InP(001) QDs \cite{Syperek2018}. For highly asymmetric confining potentials like those in InAs/InAlGaAs/InP(001) quantum dashes, the PL lifetime consists of two short and long components \cite{Gawelczyk2017}. Nevertheless, the average PL lifetime of $\SIrange{1.45}{1.8}{\nano\second}$ is comparable to the obtained here, and indicates the main influence of the exciton confinement regime on the observed PL dynamics.

\newmg{A significantly different value is obtained for family G, where the analysis yielded the PL lifetime of $2.0\pm0.2$~ns. This family shows also a different PL quenching characteristic than other families, as discussed in the next section, which may suggest a much different exciton confinement nature reflected in a much longer PL decay time.}

Finally, the observed long PL decay time of $0.9\pm0.2$~ns for the 0D WL states represented by the PL band peak A at $\sim 0.93$~eV confirms the localized nature of the states \cite{Syperek2013} since the lifetime of excitons confined in the 2D WL is expected to be significantly shorter.

\subsection{Temperature-dependent photoluminescence}
In this section, we describe temperature-dependent PL that provides information on the carrier confinement parameters in the WL and QDs and the thermal carrier redistribution processes involving 0D and 2D states in this system. 

\begin{figure*}[t!]
\includegraphics[width=\linewidth]{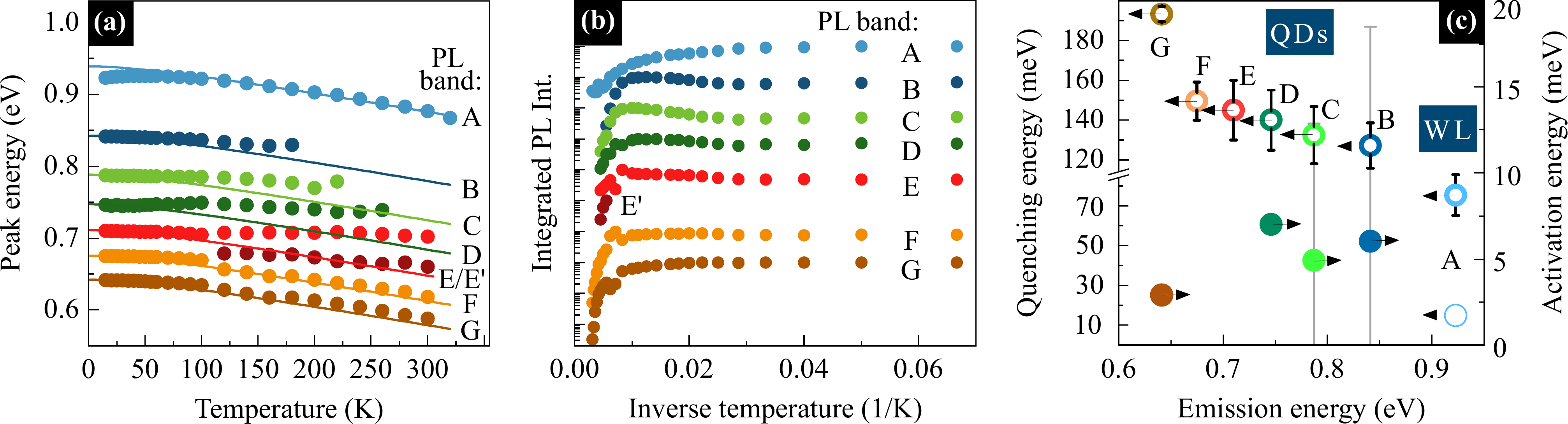}
\caption{\label{fig:Temp_analysis}(Color online) Analysis of temperature-dependent PL spectra: (a) PL band peak energy, (b) integrated PL intensity, (c) activation (closed circles) and quenching (open circles) energies for the emission bands taken from the Arrhenius fit.}
\end{figure*}
First, the analysis is focused on the thermal shift of the PL peak energy presented in \subfigref{fig:Temp_analysis}{a}. The shift is compared to the well-known Varshni relation for the temperature dependence of the energy gap for bulk InAs material \cite{Varshni1967}:
\begin{equation}
 E_{\mathrm{g}}(T) = E_{\mathrm{g}}(0)-\frac{\alpha T^2}{T+\beta},
\end{equation}
where $E_{\mathrm{g}}(0)$ is the estimated transition energy at $T=0$~K, and $\alpha=2.76\times 10^{-4}$~eV/K$^2$, $\beta=93$~K are parameters for InAs. 

\newms{From \subfigref{fig:Temp_analysis}{a}, we infer that for the PL band A (blue circles) the peak energy follows the Varshni relation (solid blue line) only at elevated temperatures. At $T<120$~K, the peak position remains nearly $T$-independent or slightly drops when $T$ decreases. The effect reminds the so-called "S-shape" function for a QW \cite{Baranowski2011, Baranowski2012}. In this case, photo-injected carriers populate 0D-like states residing below the QW's mobility edge (2D states) and the resultant emission under low photo-excitation conditions involves the localized states only. 
\newph{The WL can be considered as a QW with additional localization potentials which come from, e.g., fluctuations of its width and chemical content, as well as variations of local strain.}
However, the optical response from the 2D WL \newph{states} should be present: (i) in the PL spectrum, once the 0D states are fully occupied preventing carrier transfer process from 2D to 0D WL states, or (ii) in the $\Delta R/R$ trace, where all the 2D states with a large transition oscillator strength are accessible in the optical absorption process.
Both are observed experimentally among the registered data set.
The extrapolation of the Varshni trend to low temperatures predicts the optical response involving 2D WL states to be observed at $\sim0.94$~eV.
While the inset to \figref{fig:PL_MR} shows the shift of the PL band A towards the predicted energy range with \newph{increasing} excitation power, the $\Delta R/R$ experiment presented in \figref{fig:PL_MR} revealed the highly intensive feature at 0.94~eV, identified as the optical absorption process involving 2D WL states.
The energy distance between the PL band A and the 2D WL is $\sim10$~meV at $T=10$~K, which may serve as a rough estimate of the carrier localization energy.
These observations confirm the existence of 0D WL states able to localize carriers at low $T$ and suggest their impact on the thermal carrier redistribution process in the structure. A similar issue has been previously studied for (In,Ga)As/GaAs QDs \cite{Syperek2013}.}

\newms{The temperature dependence of QDs-related PL bands B--G is different from the one observed for the WL emission. At $T<40$~K, the peak energy follows the Varshni trend. However, with increasing $T$ the redshift is either stopped (bands B--E) or weakened (F and G). This so-called anomalous energy shift is only partially caused by the bandgap reduction of the QD material \cite{Lei2006}. It is predominantly related to the migration of carriers between different QD families and within each of them, as well as it involves exchange of carriers with the WL reservoir. During the thermally activated redistribution, carriers can be captured by QDs characterized by high emission energy. Therefore, the high-energy tails of QD size distributions in each of the families become optically active, shifting the PL-band peak energy towards higher energies, thus compensating for the effect of the thermal bandgap reduction \cite{Lei2006}.} 

The analysis of the temperature-dependent PL intensity presented in \subfigref{fig:Temp_analysis}{b} provides further information about the carrier redistribution in the investigated structure. Each data set was fitted with the Arrhenius-type formula: 
\begin{equation}
I(T) = \frac{I_0}{1+\sum_{i}B_i\exp\left( \frac{-E_i}{k_{\mr{B}}T}\right)},
\end{equation}
where $E_i$ is the average \newph{activation energy (or PL quenching energy)}, and $B_i$ is the amplitude of the activation (PL quenching) process. The respective energies are summarized in \subfigref{fig:Temp_analysis}{c}.

\newms{For the already discussed PL band A originating from the 0D WL states \newmg{overlapping with 2D WL states}, the PL intensity is thermally quenched, what is characterized by two energies of \newmg{$16\pm2$~meV and $75\pm22$~meV}, represented by open blue circles in \subfigref{fig:Temp_analysis}{c} at $\sim0.93$~eV. The lower energy is \newmg{close to} the estimated energy of additional localization in the WL most probably caused by the fluctuations in the chemical content or/and the WL width variations. \newmg{Note that these two do not have to match, as the latter is a difference between exciton levels in the WL and in an additional potential trap, while the former is composed of such a difference for just one of the carriers (most probably the hole) and exciton binding energy}. Since the PL quenching is rather weak in the low-$T$ range, most of the exciton population is trapped back to the 0D WL states and recombines there, while only small exciton population feeds QDs. \newmg{In the range of higher temperatures} more complex PL quenching processes occur. \newmg{When thermal energy is larger}, the WL reservoir is depopulated by the carrier capture in QDs, and at the same time, re-activated carriers from QDs supply the WL reservoir. \newmg{On this background, the escape of holes from the WL to the barrier seems to be the most important process, as the higher quenching-related energy corresponds well to the calculated difference in hole levels plus the exciton binding energy in the WL}.}

\newms{Prior to the discussion on the characteristic PL quenching and activation energies for QDs, it is justified to assume that in the low-$T$ limit free carrier migration is negligible. This comes from the calculated electron-hole Coulomb correlation energy between $\sim9.5$~meV for 14-ML-high QDs and $\sim 20$~meV for the flattest ones, which corresponds to $T \sim 110$--235~K. At higher temperatures, electrons and holes can escape from the confining potential with characteristic activation energies composed of the energy distance to WL states ($\delta_e$, $\delta_h$) or barrier band edges ($\Delta_e$, $\Delta_h$), in which we also include the exciton binding energy ($E_{\mathrm{b}}$), as the exciton has to dissociate.}

\begin{figure}[tb]
\includegraphics[width=\linewidth]{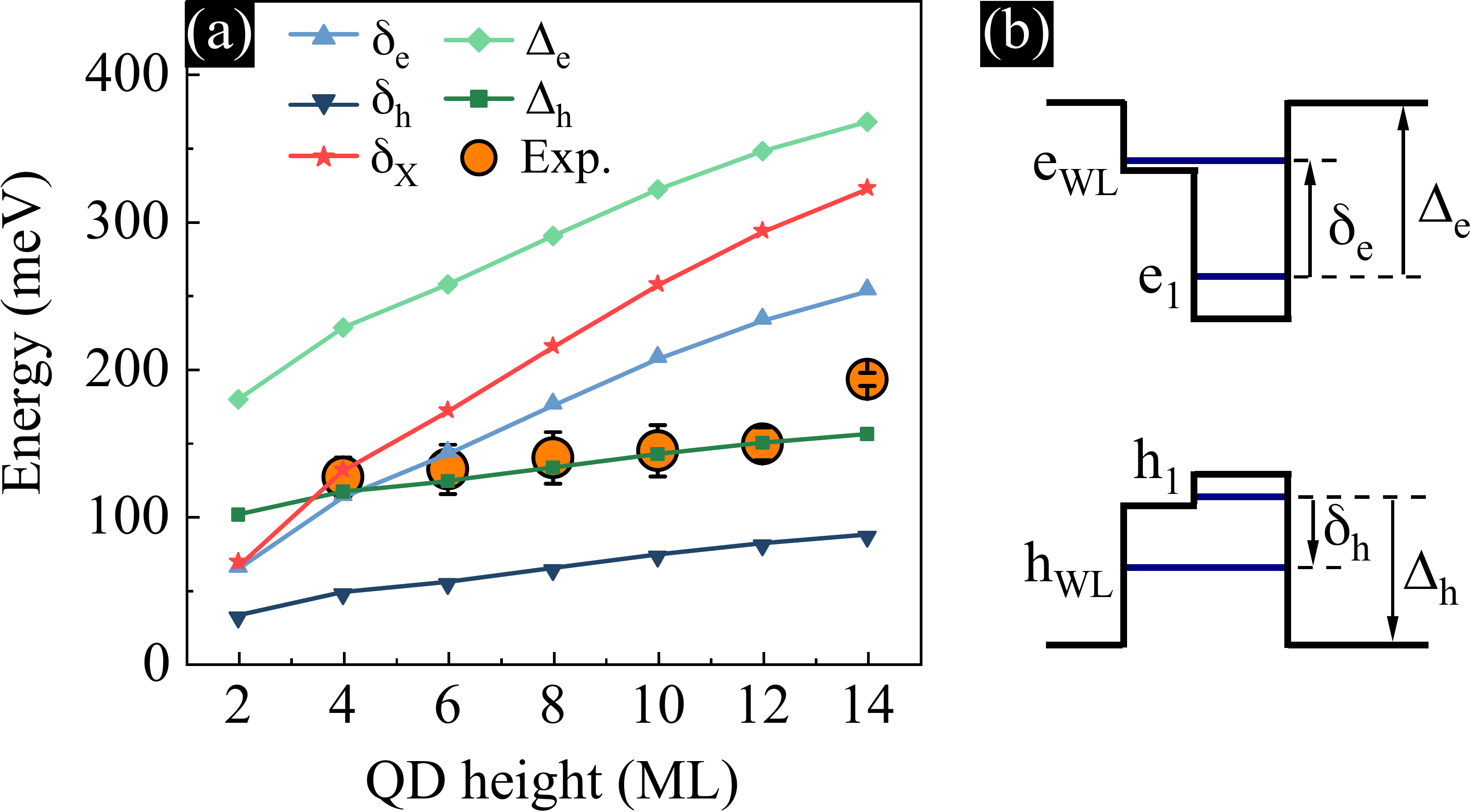}
\caption{\label{fig:Activation_energies}(Color online)
(a) Comparison between experimentally obtained PL quenching energies (orange circles) and calculated energy distances between single-particle electron $\left(e_1\right)$ and hole $\left(h_1\right)$ ground states confined in a QD and (i) 2D WL electron $\left(e_{\mathrm{WL}}\right)$ and hole $\left(h_{\mathrm{WL}}\right)$ states,
$\delta_{\mathrm{e}} = e_{\mathrm{WL}} - e_1 + E_{\mathrm{b}}$, 
$\delta_{\mathrm{h}} = h_1 - h_{\mathrm{WL}} + E_{\mathrm{b}}$, and 
(ii) the InAlGaAs barrier, where $\Delta_{\mathrm{e}}$ and $\Delta_{\mathrm{h}}$ are the distances from QD states to the the conduction and valence band edges in the barrier, respectively.
$\delta_{\mathrm{X}} = \delta_{\mathrm{e}} + \delta_{\mathrm{h}} + E_{\mathrm{b}}$ is sum of energy required to dissolve the exciton ($E_{\mathrm{b}}$) and excite both carriers to the WL ($\delta_{\mathrm{e}}$, $\delta_{\mathrm{h}}$).
(b) Energy diagram of the confinement potential for the system with a QD and the WL.}
\end{figure}
\newms{Let us now refocus on the $T$-dependent PL intensity from QDs. For families B--D, first an enhancement is observed, followed by a decrease of intensity at elevated temperatures.
The extracted activation energy connected with this enhancement is $\sim 6$~meV with variation within the uncertainty bounds, as depicted with full circles in \subfigref{fig:Temp_analysis}{c}. This is close to the determined PL quenching energy for the 0D WL states suggesting that QDs may acquire excitons released from the WL already at low $T$. However, this feeding process competes with the thermal escape of carriers at elevated temperatures. Energies connected with quenching of PL from QD families are depicted with open circles in \subfigref{fig:Temp_analysis}{c}, with values varying between $127\pm11$~meV and $150\pm10$~meV. As it could be expected, a monotonic increase of quenching-related energies with rising QD size (families B--F) is present, as the localized states are sunk deeper into a wider potential.}

\newms{To propose a scenario underlying the carrier activation process, one can compare the obtained PL quenching energies with calculated energy distances between the lowest single-particle electron ($e_1$) and hole ($h_1$) states confined in a dot and respective 2D WL ground states ($e_{\mathrm{WL}}$, $h_{\mathrm{WL}}$) or the InAlGaAs barrier ($e_{\mathrm{bulk}}$, $h_{\mathrm{bulk}}$). Results are presented in \subfigref{fig:Activation_energies}{a} together with a sketch of confining potential in a QD and the WL in \subfigref{fig:Activation_energies}{b}. Note that all the calculated energy differences $\delta_{\mathrm{e}} = e_{\mathrm{WL}} - e_1 + E_{\mathrm{b}}$, $\delta_{\mathrm{h}} = h_1 - h_{\mathrm{WL}} + E_{\mathrm{b}}$, $\Delta_{\mathrm{e}} = e_{\mathrm{bulk}} - e_1 + E_{\mathrm{b}}$, $\Delta_{\mathrm{h}} = h_1 - h_{\mathrm{bulk}} + E_{\mathrm{b}}$ include the electron-hole interaction energy, as discussed above. Additionally, with stars we also plot the calculated energy of exciton extraction to the WL, in which the difference between 0D and 2D exciton binding energies is taken into account.}

\newms{From \subfigref{fig:Activation_energies}{a}, one may notice that extraction of a single electron from a dot to the barrier cannot be responsible for the PL quenching in the considered $T$-range since the $\Delta_e$ (green diamonds) is far above the obtained PL quenching energies (red stars). From the primary PL quenching mechanism, one can also exclude hole escape from a dot to the WL. The calculated energy distance $\delta_{\mathrm{h}}$ for the dots (blue down triangles) is too small to match the PL quenching energy. Note, however, that despite relatively small energy necessary to eject a hole to the WL, the inverse process of fast hole relaxation mechanisms back to a QD at low $T$ may lead to the absence of quenching related to such energies.} 

\newms{For small QDs, with $H\leq6$~ML, at least three PL quenching mechanisms can be considered: (i) the one related to single-electron escape to the WL [blue up triangles in \subfigref{fig:Activation_energies}{a}], (ii) single-hole escape to the barrier (green squares), and (iii) whole exciton extraction to the WL (red stars). In all these cases the calculated energy distances closely or perfectly fit the respective PL quenching energies.} 

\newms{For large QDs, with $H>6$~ML, the dominant PL quenching mechanism seems to be related to single-hole escape to the barrier as the calculated energy distance $\Delta_{\mathrm{h}}$ perfectly matches the quenching energy. Since the hole extraction to the barrier is present also for small QDs, one may conclude that it can be the primary mechanism for the PL quenching in the investigated QDs.}

Finally, we comment on the observed anomalies in the temperature dependence of PL from the investigated structure. At $T>125$~K, the PL spectrum is enriched by the appearance of the new PL band labeled as E' between previously identified bands E and F.
Temperature dependence of the E' band is presented in \subfigref{fig:Temp_analysis}{a}, however, the fitting procedure did not allow for a precise determination of changes in the band intensity due to its substantial overlap with bands E and F.
The origin of the band is unknown, it could be related to another family of QDs that is split off from E and F families.
Another anomaly is related to the observation of two quenching energies for band~G.
While the higher one, $193\pm5$~meV, is in agreement with the above-mentioned general scenario for QDs, the lower one, $25\pm5$~meV, suggests the existence of another non-radiative relaxation channel that we could not identify.

\section{Summary}
In conclusion, we have performed detailed optical studies of InAs/InGaAlAs/InP(001) QDs grown by MBE employing the ripening-assisted scheme. QDs emit in the $\SIrange{1.4}{2}{\micro\meter}$ spectral range, which places them among attractive solutions for photonic applications in the near-infrared.
We have studied a multimodal emission from QDs, which reveals their grouping into at least six families according to their height.
Based on the quantitative agreement between our theoretical calculation and experimental data, we show that the dots belonging to the consecutive families differ by 2 material monolayers in QD height, contrarily to 1~ML steps typically observed for InAs/InP QDs made of binary alloys.
We suggest that this may stem from crystallographic long-range ordering in the quaternary InAlGaAs barrier alloy that stabilises the growth of InAs QDs.
\newph{Due to the minimization of the crystal lattice free energy by atomic ordering, QDs in such systems should be characterized by higher long-term stability in comparison to those embedded in alloys without ordering, what is also beneficial for future applications.}
We have also found that QDs from consecutive families differ significantly in lateral dimensions spanning the 24--65~nm range of in-plane sizes. Such geometries allowed us to calculate the relatively short PL lifetimes monotonically decreasing from $\sim1.4$~ns for flat dots to $\sim1.1$~ns for higher ones, which agrees with the mean trend from experimental data. Observation of a single-exponential decay and negligible linear polarization of PL from QDs suggest their high in-plane symmetry.
Finally, we have observed a thermal carrier redistribution process, in which localized states in the WL play the role of carrier reservoir, feeding the QDs at elevated temperatures.
\newms{Based on the comparison of calculated energy splittings between carrier states in the dots, wetting layer and in the barrier with the experimental data, we have determined that the dominant PL quenching mechanism is the escape of holes to the barrier.} 

\begin{acknowledgements}
M. G., G. S. and M. S. acknowledge the financial support from the National Science Centre (Poland) under Grant No. 2014/14/M/ST3/00821. P.~H. acknowledges financial support from the Polish budgetary funds for science in 2018-2020 via the ``Diamond Grant'' program (Ministry for Science and Higher Education, Grant No. DI 2017 011747) and from the Polish National Science Centre under the Etiuda 8 Grant no. 2020/36/T/ST5/00511.
A. M. acknowledges financial support within Preludium 8 project No. 2014/15/N/ST7/04708 of the National Science Centre in Poland. 
This work was also financially supported by the BMBF Project (Q.Link.X) and DFG (DeLiCom). 
We thank Krzysztof Gawarecki for sharing his implementation of the $\kp$ method, Janusz Andrzejewski for discussions of the calculation results, and Matusala Yacob for his assistance in the MBE growth process.
\end{acknowledgements}

\end{document}